\documentclass[12pt]{article}
\usepackage{setspace}
\usepackage{verbatim}
\usepackage{tabularx}
\usepackage{colortbl}
\usepackage{natbib}
\usepackage{hhline}

\usepackage{multirow}

\usepackage{times} 
\usepackage[top=2.5cm, bottom=2.5cm, left=2.5cm, right=2.5cm]{geometry}
\usepackage{natbib}
\usepackage{amsmath}
\usepackage{amssymb}
\usepackage{latexsym}
\usepackage{sectsty}
\usepackage{amsfonts}
\usepackage{epsfig}
\usepackage{url}
\usepackage{subfigure}
\usepackage{lineno}
\usepackage{wrapfig}

\usepackage{amsmath}
\usepackage{amssymb}
\usepackage{latexsym}
\usepackage{sectsty}
\usepackage{amsfonts}
\usepackage{epsfig}
\usepackage{url}
\usepackage{subfigure}
\newcommand{\zap}[1]{}

\usepackage{color}   

\usepackage{color}   
\newcommand{\sout}[1]{ } 

\renewcommand{\vec}[1]{{\mathbf #1}}

\begin{document}

\title{Ontology alignment: A Content-Based Bayesian Approach.}
\date{\today}
\author{Vladimir Menkov \& Paul Kantor \\ DIMACS, Rutgers University  \\
  vladimir.menkov@rutgers.edu \ \ \ \ \
     paul.kantor@rutgers.edu}

\maketitle

\begin{abstract}

There are many legacy databases, and related stores of information that are maintained by distinct organizations, and there are other organizations that would like to be able to access and use those disparate sources. Among the examples of current interest are such things as emergency room records, of interest in tracking and interdicting illicit drugs, or social media public posts that indicate preparation and intention for a mass shooting incident. In most cases, this information is discovered too late to be useful. While agencies responsible for coordination are aware of the potential value of contemporaneous access to new data, the costs of establishing a connection are prohibitive. The problem grown even worse with the proliferation of ``hash-tagging,'' which permits new labels and ontological relations to spring up overnight. While research interest has waned, the need for powerful and inexpensive tools enabling prompt access to multiple sources has grown ever more pressing. 
This paper describes techniques for computing alignment matrix
coefficients, which relate the fields or content of one database to those of another, using the Bayesian Ontology Alignment tool (BOA). 
 Particular attention is given to formulas that have an
easy-to-understand meaning when all cells of the data
sources containing values from some small set. These formulas 
can be expressed in terms of probability estimates. The estimates themselves are given by a ``black
box'' polytomous logistic regression model (PLRM), and thus can be easily  generalized to the case of any arbitrary
probability-generating model.  The specific PLRM model used in this example is the BOXER Bayesian Extensible Online Regression model.
\end{abstract}

\section{Background and Problem}

In the early part of this century there was a surge of interest in a technical problem called ``ontology alignment''. In this context, the word ``ontology'' refers to the system for naming or labeling real world entities. Although that has some similarity to the much older use of the word ``taxonomy'' to describe the idea, the more recent researchers really sought to unite the computer science challenges with the deeper philosophical questions of the nature of the concepts themselves. A review of the literature shows that this topic is less active currently, although there are no plausible claims that it has been satisfactorily solved. The literature continues to grow, and much of it can be accessed through the excellent resource maintained by Pavel Schvaiko at \url{http://www.ontologymatching.org/publications.html}.

As the quantity of machine readable information increases at an incredible pace, driven both by the ``Internet of things'' and by the fact that everyone with an Internet connection can become an author, the problem of ontology alignment, while unsolved, has become significantly more important. In the commercial sector the problem is closely related to something called ``database alignment.'' A typical setting for that problem is a large corporation, each of whose operating units has maintained databases for purposes of billing, customer service, maintenance, etc. As part of the global computerization of business, these companies faced a need to align their databases. Consulting firms developed effective, albeit quite expensive processes for this database alignment. Those processes typically involved direct discussions with the owners and maintainers of each of the databases, in order to map them to a single shared ontology, or view of the world. Once that map had been completed (which, although the challenge might appear philosophical, was greatly facilitated by reference to the common real-world operating picture) then the mappings could be programmed, and each database became accessible to all the others, through a common query language.

In other fields, typically represented by governmental operations, this path is enormously difficult to follow. We need only point to the hugely expensive and ultimately abandoned effort by the Federal Bureau of Investigation to develop common computer systems.  Those difficulties, in 2005, were described bitterly as:

\begin{quote}
Sen. Leahy offered another, more whimsical analogy for Trilogy: the 1993 movie Groundhog Day, in which Bill Murray wakes up each morning to relive the same day. Since 1997, proposals for modernizing the FBI's technology and processes have emerged again and again, culminating with Trilogy. Trilogy itself then underwent a cyclic series of evaluations and funding requests until Congress finally learned that its third leg, VCF [Virtual Case File] , might never materialize.  [cite info world article: tag BOA].\footnote{  https://www.infoworld.com/article/2672020/anatomy-of-an-it-disaster--how-the-fbi-blew-it.html}
\end{quote}

Trilogy, at a cost of close to half a billion dollars, was simply never completed. Today, there are literally thousands of agencies that generate or hold information whose coordinated use could address some of the most serious problems faced by the nation. 

To give just two prominent examples, we consider the opioid crisis, and the problem of domestic mass assaults. Information relevant to the opioid crisis is in the hands of law enforcement, of healthcare services, and of the public-facing social media. Overdoses and requests for help appear in the health service records. Unlawful behavior related to addiction appears in the records of law enforcement. Warning signs of a growing dependence, intention to use, or intention to commit crimes may all appear in social media. 

Information related to mass assaults ordinarily is not in the data stores of law enforcement until it is too late. There are however relevant government records which could or do exist and could be coordinated. As an example, the Las Vegas shooter had assembled a very large array of weapons and ammunition, through legal purchases. That information was not available to the hotel when he checked in, and might conceivably have been used to examine his luggage and prevent the tragedy. We learn, after the fact, that there are often social media traces of the growing intent to commit massive harm. Although naive combination would lead to a great many false alarms, and would raise legitimate concerns about governmental invasion of privacy, nonetheless one can imagine and hope for a proper and constitutional path towards the integration of these warning signs in time to remove the weapons from the perpetrators. There are complex social and political issues involved here, and those will not be addressed in this paper

The two examples above serve to illustrate the use case considered here. While the agencies holding information might be quite willing to cooperate, the labels that they use to describe the information will be drastically different. Because of this, the process of aligning  the data sets, or aligning  the ontologies, can become difficult and costly. Nearly all agencies cannot afford to hire the expert consultants and spend the time needed to accomplish this matching. In fact, they face something of a ``chicken-and-egg'' problem: They cannot know whether a particular data source will add enough value to justify the costs of aligning it until they have already spent those costs.

In each of these cases, it is likely that social media will provide an important source of very contemporary information. On the face of it, these are not data stores with fields and entries. However, we can reconceptualize them in that way by considering that hashtags represent a kind of ontology. It is poorly structured (if at all) and can rapidly change. But the collection of postings labeled by a particular hashtag can be considered to represent a conceptual class, in much the same way that locations or organizations are classes. Because these change rapidly, a phenomenon that has been called ``folksonomy,'' a combination of ``folk'' and ``taxonomy,'' it is particularly important to be able to update the relations among them, in order to follow trends in labeling and tagging. \footnote{We are not proposing the kind of intensive monitoring of social media that is a feature of daily life in China today.} All of what we present here, can be immediately adapted, {\it mutatis mutandis} to the alignment of hashtags, and similar folksonomies. For a review of the status of folksonomy research as of 2007, see, for example, (\cite{laniado2007using}). For an example of the possibilities of transforming a folksonomy into a hierarchical structure see also (\cite{almoqhim2013approach}). 

For the two examples given here, opioid addiction, and mass assaults, none of the  cooperating agencies belong to  any single organization which might undertake to bring them together and cover the expense. On the other hand, if agencies are willing to share the contents of their databases, then they will (logically, at least) be willing to process those databases, inexpensively, to provide the kinds of metadata that will make alignment easy. The key to this process involves two steps: selecting the metadata, and calculating the probability of various possible alignments.

Ordinarily, one might think of the metadata as being information about the ownership, size, formats, etc. of the database. Our analysis recognizes that when the challenge is to align the fields or entities of the database,  what we need is the metadata characterizing each of those fields or entities. Previous research, and there has been a great deal of it, as reviewed NEED CITE, focuses on two kinds of metadata: the literal names of the entities, and the network of relations among the entities. Our approach is at the same time both simpler and more complicated. It is simpler, because we do not deal in the abstract world of the ontology of entities and concepts itself. It is more complicated, because we exploit access to the data fields to develop characterizations of each of them separately. 

These characterizations are inspired by the very substantial progress that has been made over the last 30 years, in the automatic indexing of textual material. The successful search engines often reduce a document to nothing more than a set of numbers characterizing the frequency with which words appear in the documents, a so-called vector  representation. While it is easy to generate examples for which this representation does not work, the enormous success of search engines and recommendation engines shows that it ``works well enough.''

There are other ways to transform texts into vectors. Another important representation, which has been shown useful in retrieving corrupted texts (\cite{kantor2000trec}), and is said to have been useful in retrieval of texts in unknown languages (\cite{damashek1995gauging}), is a representation by the frequency of strings of consecutive characters. These are sometimes called ``shingles'', or ``n-grams.'' The approach presented here can be thought of as representing the entire content of a specific field in a database as a structured text. Related research has been described by (\cite{chowdhury2002collection}), and appears to be incorporated in some commercial products for database alignment.

The contents of a given field (or ``column,'' if the database is thought of as a single table)  is broken into natural units which we will call cells. Cells contain strings of characters. A first representation of the collection of cells is the distribution of the lengths of those strings. If all or nearly all of the strings have the same length, then we are surely dealing with either numerical identifiers, or terms in a local artificial language, used to describe the contents of that field. We can then look at the distribution of characters in the field. In the first case numbers and some punctuation symbols will appear, in the second case the characters could be numbers but are more likely to be alphanumeric strings from a controlled vocabulary. Similar reasoning, as be detailed in this paper, can carry the matching to higher levels of precision.

The task can be thought of as a problem in Artificial Intelligence, Machine Learning or Classification. In fact these disciplines overlap in problems of the kind addressed here.  Automatic determination of the distinguishing characteristics of
any {\it class} can be viewed as a machine learning problem. Thus our approach represents a (less common)  application of classification to the  problem  generally called ``ontology alignment.''  Here the word ontology refers to the kind of naming structure that used to be called a thesaurus or dictionary of relations among terms.  In other words, the kinds of information associated with specific locations in the ontology is assumed to define a class. The problem of interest is to learn a characterization of the kinds of entities that belong to that class. In particular, all of the entities with which we deal in this problem are strings, and they designate corresponding entities in the physical world, for example:

\begin{verbatim}
name_last: Smtih;
name_first: Sarah;
PhoneMobile: 001-23-452-234243
\end{verbatim}

The entity of interest to us is, for each characteristic of the physical entity, the string representing it. Thus the strings of interest are: \begin{verbatim}"Smtih", "Sarah", "001-23-452-234243" \end{verbatim}

We  consider the situation in which we have access to a number of data sets that correspond in whole or in part  to a given ontology (or, more precisely, to a given selection of characteristics of some  entities or events). In addition, the entries in these data sets are all expressed in a common natural language. For example, one set of entries may correspond to a ``viewer's response to a movie'' and might contain entries such as {boring, splendid, thrilling, heart-breaking}. It would not contain entries such as {Coppola, Ford, Huston}, which would be found, for example, in entries corresponding to the (names of) directors of movies. 

**** \\
Similarly, we may consider working with several databases in which related information is stored, but using different ontological labels as shown in Table \ref{tab:paulVersions}. 

\begin{table}[h!]
\begin{center}
 \begin{tabular}{|| l| l| l| l||} 
 \hline
 Database & FieldName & Content & Comment \\ [0.5ex] 
 \hline\hline
DB1 & Client & Paul B Kantor & full name natural \\ 
DB2 & Flyer & Kantor & just last name \\
DB2 & FlyerFB & PaulB & first name with MI \\
 DB3 & AccountHolder & Kantor, Paul B.& Full name, last first \\
 DB4 & Patient & Kantor & Last alone \\
 DB4 & Patient\_FN & Paul & first alone \\
 DB4 & Patient\_MN & B & middle alone \\ 
 [1ex] 
 \hline
\end{tabular}
\caption{Alternate Version of Related Information}
\label{tab:paulVersions}
\end{center}
\end{table}

We propose that the relations among entities can be partially discerned through ``alignment by content analysis.'' Before discussing the technical details, we consider several example cases.

{\bf Case 1}. The case of minimal information. Suppose that an investigative unit has come into possession of a gigabyte hard drive containing a great many  structured data objects. (Imagine the results of a successful raid on a drug headquarters, or the seizure of documents in an investigation of suspected fraud.)
These might be in the form of Access tables, or MySql tables, or tables from Oracle or Sybase. Alternatively, they might be tables that are presented as (and manipulated within) a spreadsheet program such as Microsoft Excel (.xls, .xlsx, etc.), or Open Office Calc (.odf). They might be presented as tables in an Adobe (.pdf) document, whose layout and structure make it obvious to a human reader that they represent structured data. [See Figure 1]. Other cases, which are again obvious to a human reader, but pose challenges for machine processing, include varied format presentations of the same information in paragraph style. 

The investigative unit needs to know what kind of information is contained in each of the fields of these structures, in order to align them with other information gathered from other sources. In the most formal cases the fields will have been given names in the data structure. But even in those cases the analyst or data entry clerk may have simply left default names such as -- Field\_1, Field\_2, or Variable\_1, Variable\_2 -- since, to the users, the meaning of each field is transparent. Or the names might vary in ways that seem natural to a human reader, as in Table \ref{tab:paulVersions}, but show few commonalities that will be obvious to a computer.

{\bf Case 2.} Several agencies of Federal, State, Local, or Tribal Law Enforcement US Government seek to collaboratively share information, but are prohibited by law from sending wild card queries to each other, and may not even be able to directly share the details of the ontology associated with a particular data collection. However, it might be legal  to share a profile or model of the contents of specific fields, as described below, while not revealing any personally identifiable information. If this process can be made so inexpensive as to be within budget for small agencies, it would  facilitate  effective construction of inter-agency inquiries and rapid data-sharing.

{\bf Case 3.} A large organization with thousands or millions of legacy databases, and no legal barriers to inquiry and response may wish to align the databases using techniques that minimize the requirement for human labor.

In all three of these cases, we anticipate that there must  be a combination of machine and human intelligence in the actual  solution. The algorithms presented here can work to produce {\it candidate} alignments of fields which can then be edited and/or finalized by a human operator {\em for the cases of interest} using tools such as, for example, AgreementMaker
(\cite{Cruz}).

\section{Algorithm and Approach}

Our approach builds on the idea that one can model the contents of a field, treating the information provided as a multi-class (``polytomous'') classification problem.  There is a  general class of applicable tools are called PLRMs (polytomous logistic regression models) and we illustrate the approach using a specific tool, BOXER, which is available at (\cite{BOXER}). Some experiments to assess the efficacy of BOXER as a general learning tool are reported in (\cite{BOXER-TR1}). That report also contains a detailed discussion of the algorithm itself, and the theoretical underpinnings of the alternative approaches that it offers.

\subsection{A multi-level approach}\label{sec:levels}

The concept of classification can be applied at four distinct levels. Two of these have to do with the representation of the contents of the cells that make up a particular field. The other two have to do with the representation of the field as the collection of its contents.  This can be thought of as the distinction between individuals and populations.

Representation of individuals, when those individuals are text strings (or unicode strings) may be done on the basis of ``words'' (for example for European languages), or on the basis of  ``character n-grams'' which applies equally well to texts lacking the notion of a word separator.  We will report primarily on n-gram representations.

Representation of a whole population, on the other hand, can exploit characteristics that are of little help in classifying individuals. For example, the distribution of lengths of the entries may be characteristic of the field as a whole, but have little discriminating value for an individual entry.  In addition, the frequency with which an entry is missing (or takes the value NUL) can be characteristic of a field in an  ontology, as we shall see below.

Another way of thinking of the problem is this: do we wish to classify each individual entry appearing in an unmatched field, to the closest known field? Or do we wish to classify the whole field to which this entry belongs, to the closest match?  While ordinary classification seeks to label the individual cases, the ontology alignment problem has the important additional information that a correct alignment will apply to all of the entries in the newly aligned field. We will have occasion to exploit this principle, which we call the {\bf principle of uniform class}, in what follows, with considerable effect.

There are, however, caveats to the principle of uniform class, as suggested by Table \ref{tab:paulVersions}. One ontology may contain full names, while another presents family names and given names separately. Both of the latter match to the former. On the other hand, if the names have been given separately in the labeled ontology, the unlabeled ontology will, to some degree, match to both of the parts of names. While it can be asserted that ``the assignment will be equally strong for all entries in the full name field'' we recognize that this is a partial solution. A better solution would recognize that the full name can be parsed into pieces that exhibit better alignment with the name parts in the given ontology.

An additional caveat flows from the fact that human error is universal, and data entry checking code is only as good as the programmer who wrote it. In fact, when the structured data is extracted, by scanning and OCR from non-structured documents, there was no error checking at the point of creation and all kinds of errors may occur. That is why probabilistic methods are called for, and are most effective.

We have not found it easy to identify publicly available data sets that are challenging, and that do not contain information about United States persons. Since this work is supported by the US Government, we did not feel comfortable working with information of that type.  In this note we report preliminary exploration of the several principles sketched above, using a single split data set.

\section{Detailed Formoulation of the Ontology Alignment Task}

We are given a (``labeled'') table with $N_1$ rows
and $M_1$ columns ($C_1$, $C_2$, \dots, $C_{M_1}$), representing a
sampling of data from ``Data Source'' DS1. This might be, for example, an
SQL
database table. Each cell of the table contains an entry corresponding
to some attribute of an object in the set being described. The entry itself
is drawn from some
set $\cal{V}$,
 whose  exact nature is not important, as the analysis depends only on the
 representation chosen.
Each row of
the table represents a structured data ``record'' of some kind, representing some
entity, for
example a news article, and each cell of the table corresponds to a
particular data element of the record's data item - e.g., the text
strings containing the title, the main text, the name of the first
author, the name of the second author (if any), the date, the place,
etc., of the article.

There is also another table, with $N_1$ rows and $M_1$ columns
($C'_1$, $C'_2$, \dots, $C'_{M_2}$), which represents a sampling of
data from another ``Data Source'' DS2. The entries in this table
also belong to the set $\cal{V}$.

We use the notation $z_{ij}$ and $z'_{ij}$ for the values in the cell
in row $i$, column $j$, of DS1 or DS2 respectively.

While the second table is quite different from the first, it is proposed
that the data in the two tables correspond to real-world
objects of the same, or related types. We also suspect that the
data in the records is divided into fields (columns) for the
representations in the two tables, even though the names of the columns
in the two tables may be quite different, or simply missing. [Note, for example, that if the fields are
names V1, v2, V$N_1$, for example, the name is meaningless.] Our task,
which can be called
``aligning the
ontologies,'' is to figure out which columns of the second table
correspond to which columns of the first table. We propose to
represent the results as a matrix. If the true alignment is known,
this can be considered a  {\bf ``confusion matrix.''} In any case,
the computation should yield a
specific number for each pair ($C_i$, $C'_j$). Rows of the matrix
will correspond to columns of DS2, and columns of the alignment
matrix, to columns of DS1, the ``labeled set.''

We may consider that the set of possible entries $\cal{V}$ to be represented in
 some finite-dimensional linear space $\cal{U}$, although in
a special case later in this report that fact will be of
limited importance. A specific example is the use of character 2-grams, under which the string ``banana" corresponds to the set of 2-grams \{ba an na an na \} or the vector $(1,2,2)$ in the basis $(``ba", ``an", ``na")$.

\section{Algorithm Overview}
\label{sec:overview}

The family of ontology alignment algorithms we consider is based on
underlying algorithms that can classify elements of the set
$\cal{V}$ (i.e., the set of all entries in distinct fields) with
respect to their propensity to be properly situated in various columns
of DS1. In other words, we think of ``appearing in column $Ck$" as
characterizing a class $C_k$ of entries, contained in the global set
$\cal(V)$.  Based on the characterization (representation) of the
individual entries appearing in DS2, we compute the degree of fit of
each such entry to each of the multiple classes defined by the fields
of DS1. Finally, we must compute a plausible ``assignment matrix''
whose entries link the fields of DS2 with those of DS1.  The elements
of this matrix have some similarity in meaning to similarities
computed using Language Models, as they are related to the probability
that the given ensemble of entries from a field of DS2 would have been
``generated by the model process that has given rise to the observed
collection of entries in the class $C_k$ corresponding to the $k-th$
field of DS1.''

The overall algorithmic framework can be described as follows:

\begin{enumerate}
\item[1] Consider the set of $M_1$ fields of DS1 as a single
  discrimination (that is, a set of labels) with $M_1$ distinct classes (one per field)

\item[2] Create a set of  $M_1 \cdot N_1$ training examples, each example being
  the entry in one field of one record from DS1, and carrying a
  class label representing the field in question. If the fields are labeled by meaningful strings, this class label can be the corresponding string. (When the data are
  represented with records as rows and fields as columns, each
  ``example" introduced at this step will correspond to the contents of
  one cell of this table)\footnote{It is interesting to consider the possibility of defining a similarity between class models, which would correspond to measuring the difficulty of resolving the classes. This might be done at the model level, considering models to be elements of a vector space dual to the space of representations. Alternatively, it could be assessed by applying the models to actual data. In the limit where the sample of available data is large enough, these two measures should converge.}.

\item[3] Tokenize etc. each ``example," representing it as an object in the
domain to which the classification algorithm can be applied. In the present note, this representation will be  a vector in some linear space (called a ``feature vector") and the examples will be entries from the fields of the unclassified data set DS2.

\item[4] Apply some kind of learning algorithm to that set of $M_1
  \cdot N_1$ ``examples,'' to induce a classifier model that
  probabilistically associates each ``example'' with a ``class'' (i.e., a
  field).  In this note we use algorithms implemented in the  BOXER learner toolkit. We expect that predictions for individual examples may be of poor
  quality (e.g., when several columns are all filled with ``Yes" and
  ``No" values), this modeling can, for example, only capture the fact that the
  ratio of ``Yes" and ``No" differs across fields. The result is a polytomous classifier that recognizes $M_1$ distinguishable classes.

\item[5] Create $M_2\cdot N_2$ test examples, from the entries of DS2, as in
Step 2. Convert each example to a feature vector, as in
Step 3.

\item[6] Apply the classifier model obtained in Step 4 to these $M_2
  \cdot N_2$ test examples (entries from the cells of DS2). For each one, an array of
  $M_1$ assignment values (summing to 1.0) will be produced,
  describing the likelihood that this particular entry belongs to
  the class defined by each of the columns of the training data set DB1.

\item[7] For each column $C'_j$ of the DS2 we now have $N_2$ arrays
  (one for each cell) of $M_2$ assignment values each. We then compute
  each assignment matrix value $f_{ij}$, describing the level of
  ``connectedness'' of $C'_j$ with DS1's column $C_i$, based on the
  $N_2$ values obtained in Step 6 for the cells of $C'_j$. The process
  whereby this aggregate value $f_{ji}$ is determined is not specified at this time;
  note however, that there is, generally, no guarantee that $\sum_i {f_{ij}}=1$, or that the entries in the assignment matrix
  $f_{ij}$ can be considered as conditional probabilities, something like
  $P(C_i|C'_j)$.

\item[8] While the values computed in Step 7 may or may not be
  interpreted as probabilities, the expectation is that, for each specific class arising from DB2:
  $C'_j$, a greater value of $f_ij$, as a function of $i$, the class label, corresponds to a greater degree of
  connectedness. If a single assignment is required, we can thus select
  $i ^{*}(j)=\operatorname*{arg\,max}_k f_{kj}$
   so that  $f_{i^{*}j} \ge f_{kj}$ for   any $k\ne i^{*}$. We will then claim that $C'_j$ has the closest association with   $C_i*$. In other words, we will call $C_i^{*}$ the ``best match'' for   $C'_j$.

  \item[9] If there is a human in the loop, or a succeeding algorithm that can handle ambiguity gracefully, the BOA process might be used to provide a ranked list of a few alternatives, and to transmit the values in the assignment matrix to the successor algorithms.

\end{enumerate}

\section{Some Properties of the Bayesian Model}

Let us assume that the learning algorithm used in Step 4 is efficient
enough, and is able to construct a PLRM model very close, in terms of
log-likelihood, to the optimal model for this problem. What can
be said about this model? The elements of the alignment matrix
$f(C_i|v)$
matching the various elements of $\cal{V}$, with respect to various
columns of DS1, will depend on how the elements of $\cal{V}$ have been
converted to feature vectors. However, under a certain - sometimes
reasonable
- assumption, the particular feature selection and the 
 particular linear regression algorithm may not  matter much.

{\bf Assumption 1.} The elements of $\cal{V_1} \subset \cal{V}$, the
set of all values of cells of DS1, have been converted to linearly
independent vectors.

The above assumption holds, for example, if each distinct field or class has at least one unique ``shibboleth'' - a word that occurs in no other cell whose
entire text is different from this cell's text.

This is the case, for example, if each cell contains a single word, or
is empty. Our feature space consists of all words occurring in the
cells, plus the special ``empty'' token, and each cell's content is converted to a vector with a single
co-ordinate set to the value $1$.

Under Assumption 1, the following holds about the optimal Bayesian model
that one can build:

Let $\alpha_i(v)$ be the fraction of the cells in column $C_i$ that
contain the value $v$. (Thus, $\sum_{v\in \cal{V}} \alpha_i(v) =
  1$). Then the Bayesian probability of assigning the value $v$ to
  column $C_i$ is
\begin{equation}
\label{eq:bp}
P(C_i|v) = \frac{\alpha_i(v)}{\sum_{j=1,\ldots,M_1} \alpha_j(v)}.
\end{equation}
In other words, the probability of assigning a given value $v$ to a
  particular column $C_i$ is proportional to the share of the cells
  with $v$ in the entire table that are located in column $C_i$.

\section{The Bayesian Ontology Alignment Tool}
As  discussed in Sections \ref{sec:levels} and \ref{sec:overview}, there are, broadly, three steps involved in converting the content-based alignment problem into a problem suitable for PLRM analysis:
\begin{enumerate}
\item select a representation of the contents and the field itself;
\item select a method for computing the alignment value for each entry in a field (or the ensemble of entries in the field) in the new data set with respect to a given field of the given ontology;
\item select a method for aggregating these alignment values, computed for individual entries, into a measure of fit for the field as a whole.
\end{enumerate}

Building upon the BOXER (the Bayesian Online EXtensible Regression
toolkit), we have created a software application, named the Bayesian
Ontology Alignment tool (BOA), which allows the user to mix and match
techniques used for the three steps of the matching process. Its
operations can be controlled via three groups of command-line options,
responsible for, respectively, representation (``tokenization'') of
the data; ``classification'' of the cells, i.e., matching their
conents to the fields; and ``aggregating'' the values so obtained, to
construct the final alignment matrix.

The ``tokenization'' options support the feature-vector representation
of the text of each cell using features of a specified type(s): words
and/or $n$-grams of a specified length.

The ``classification'' options support a number of techniques for
finding a more or less approximately optimal polytomous Bayesian model in
terms of the log-likelihood, as well as in terms of regularized
(penalized) log-likelihood.  The $k$-NN classifier is also supported.

Sections \ref{sec:agg} and \ref{sec:unequal} will
discuss options available to us, and implemented in BOA, for
``aggregating'' the individual cells' probability values into the
alignment matrix values for the fields of the two data sources.

When the three-step scheme outlined above is used, it is possible to
treat one data source as the ``training'' set, and the other as the
``test set.'' The fields and cells of the first set are used as,
respectively, classes and examples based on which a classifier is
constructed, and then that classifier is applied to cells of the
second data source.

In addition to this main approach, and in a more experimental way, two
``symmetric'' techniques, described in Sections \ref{sec:sym1} and
\ref{sec:sym2}, are implemented in BOA as well. These approaches
involve both data sources in a symmetric way, so that it is not
possible anymore to refer to them as ``training'' and ``test set.''

The distinction between the methods is reflected in the structure of
the application's command line: when using a three-step method, one
needs to indicate the two data sources with the {\tt train} and {\tt
test} commands, while for the symmetric approaches, special two-argument commands are used.

More detailed documentation on the BOA application is available at the following URL:\newline
\textcolor{red}{ http://bit.ly/qst04u  Does not work}


\section{Formulas for Aggregating Assignment Values}
\label{sec:agg}

{\em (These three approaches are available in BOA with the usual combination of the  {\tt train} and  {\tt test} commands.)}

{\bf Assumption 2.} Every entry found in DS2 is also found somewhere in DS1.

In other words, $\cal{V}_2 \subset \cal{V}_1$, where $\cal{V}_2$ is
the set of values of cells of DS2.

This is generally ``not'' the case - and when it is not the case,
tokenization and conversion from $\cal{V}$ to the feature space {\em
  do} matter; but we can gain some insight from considering this
  special situation. This situation will occur
  when there is a limited ``controlled vocabulary" and a relatively large
  training set. In the extreme case the vocabulary might consist of {YES, NO}
  and all fields contain
essentially the same values, albeit in different proportions.

Similarly to the definition of $\alpha_I(v)$, let us define
$\gamma_j(v)$ as the proportion of the entries in DS2's column $C'_j$
that have the value $v$. Thus,  $\sum_{v\in \cal{V}} \gamma_i(v)=1$.

Let us consider how, in Step 7 above, individual proportions for cells
within a column can be aggregated into the assignment matrix relating
the columns.  Note that an ideal computation would be one that yields
the probability that column $j$ of DS2 should be assigned to column
$i$ of DS1.

{\bf Arithmetic mean.}
One way to compute the alignment matrix value $f_{ij}$ is to average the $f(C_i|v)$  for all cells of the column $C'_j$, which yields the averages
\begin{equation}
\label{eq:R-def1}
R(C_i,C'_j) \equiv  \frac{1}{N_2} \sum_{k=1}^{N_2} P(C_i|z'_{kj}).
\end{equation}
We can directly use these averages $R(C_i,C'_j)$ as the assignment matrix elements linking columns of DS1 and DS2:
\begin{equation}
\label{eq:ag-ar}
f_{ij}^{\rm method 1} = R(C_i,C'_j)  =  \frac{1}{N_2} \sum_{k=1}^{N_2} P(C_i|z'_{kj}).
\end{equation}
Under Assumptions 1 and 2, they would become
\begin{equation}
\label{eq:ag-ar2}
f_{ij}^{\rm method 1} = R(C_i,C'_j)  = \sum_{\cal{V}} \gamma_j(v) P(C_i|v)
\end{equation}

An advantage of this method is that $\sum_{i=1\ldots,M_1} f_{ij} = 1$, and the values in the assignment matrix can be easily interpreted as probabilities. Moreover, if Assumption 1 holds, the alignment matrix would be symmetric when the two data sources are identical (i.e., $\gamma_i(v)=\alpha_i(v)$ for all $i$), since in this case
$$
R(C_i,C_j) =\frac{\sum_{\cal{V}} \alpha_i(v)  \alpha_j(v)}{\sum_{j=1,\ldots,M_1} \alpha_j(v)}.
$$

{\bf Geometric  mean.}
One may use the geometric mean instead of the arithmetic
mean, computing the alignment matrix coefficients as
\begin{equation}
\label{eq:ag-geo}
f_{ij}^{\rm method 2} =
 \left(\prod_{k=1}^{N_2} P(C_i|z'_{kj})\right)^{ \frac{1}{N_2}}=
\prod_{\cal{V}}  P(C_i|v)^{\gamma_j(v)}.
\end{equation}
Multiplying probabilities can, of course, be interpreted as adding their logarithms.

Since the geometric mean of non-negative numbers is never greater than their arithmetic mean, we know that $f_{ij}^{\rm method 2} \le f_{ij}^{\rm method 1}$ for all pairs of columns, and the values for a given $j$ will no longer sum to 1.

We note also that the geometric mean is zero when any of the
participant columns is zero. Thus if even a single cell of the column
$C'_j$ contains a value that is {\em not} found in the column $C_i$ of
DS1, then $f_{ij}^{\rm method 2}$ will be 0. If the cells of column
$C'_j$ mixes values in a way not seen in {\em any} column of DS1 ---
that is, for every $i \in {1, \ldots, M_1}$ there is some value $v$
found in $C'_j$ but not found in $C_i$ --- then {\em every} coefficient
$f_{ij}^{\rm method 2}$ for column $C'_j$ will be zero; that is, Method 2 (eq. (5)) would
conclude that $C'_j$ is not similar at all to any column of DS1. This seems a fatal flaw. In similar situations models may add a small positive number to prevent ``Bayesian annihilation.'' The resulting model becomes:
\begin{equation}
\label{eq:ag-geo2}
f_{ij}^{\rm method 3} =
 \left(\prod_{k=1}^{N_2} P(C_i|z'_{kj})+\epsilon \right)^{ \frac{1}{N_2}}=
\prod_{\cal{V}} \left( P(C_i|v)+\epsilon \right) ^{\gamma_j(v)} .
\end{equation}

{\bf Cosine similarity.}  Both of the methods above are not
particularly good when what we want to distinguish are columns that
are composed of the same values and are only different by the
proportions of those values. What we'd like to have is an assignment
matrix whose element $f_{ij}$ is maximized whenever the vector
$\vec{\gamma_j}$, whose components are the relative frequencies
$\{\gamma_j(v)\}_{v\in \cal{V}}$ of various values in $C'_j$ is the
same as the vector $\vec{\alpha_i}$ of relative frequencies of various
values in $C_i$.  A natural approach here would be a weighted cosine formula,
$$
f_{ij}^{\rm cosine \ method} = \frac{\sum_v{ \alpha_i(v) \gamma_j(v) \phi(v)}}{
\left(\sum_{v\in \cal{V}}{\alpha_i(v)^2  \phi(v)}\right)^{1/2}
\left(\sum_{v\in \cal{V}}{\gamma_j(v)^2  \phi(v)}\right)^{1/2}
},
$$
with some reasonable term-weight function $\phi(v)$.

But we would like the formula for $f_{ij}^{\rm cosine \ method}$
to be computable purely on the basis of alignment values for the cells,
$f(C_i|z'_{kj})$, and without explicitly using the values of
$\alpha_i$ and $\gamma_j$. This would allow us to naturally expand the
use of the formula even on the situation when Assumption 1 does not
entirely hold.

Considering the formula for the Bayesian probability (\ref{eq:bp}), we
note that the following weight would work very well for our purpose:
$$
\phi(v) = \frac{1}{\sum_{j=1,\ldots,M_1} \alpha_j(v)}.
$$

This kind of weight is readily interpreted as the inverse of the
overall frequency of a particular cell value in the entire table
DS1, and, like the ``idf" factor of information retrieval, it seeks to emphasize the more informative (that is, less common) feature values. This leads to
the following scoring formula:
\begin{eqnarray}
\label{eq:ag-cos2}
f_{ij}^{\rm cosine \ method} &=& \frac{\sum_v{ P(C_i|v) \gamma_j(v)}}{
\left(\sum_{v\in \cal{V}}{P(C_i|v)\alpha_i(v)}\right)^{1/2}
\left(\sum_{v\in \cal{V}}{\gamma_j(v)^2  \phi(v)}\right)^{1/2}} \\
&=& \frac{ R(C_i,C'_j) }{ R(C_i,C_i)^{1/2}
\left(\sum_{v\in \cal{V}}{\gamma_j(v)^2  \phi(v)}\right)^{1/2}}.
\end{eqnarray}

The values $R(C_i,C'_j)$ and $R(C_i,C_i)$ are precisely the arithmetic
means introduced in eq. (\ref{eq:R-def1}) above, and are computable without
reference to Assumption 1. However, the last factor,
$\|\vec{\gamma_j}\|_\phi=\left(\sum_{v\in \cal{V}}{\gamma_j(v)^2 \phi(v)}\right)^{1/2}$ is
  {\em not} computable without reference to Assumption 1. But we note that
  the expression  $\|\vec{\gamma_j}\|_\phi$ is a factor common to $f_{ij}$ for all $i$ for a given
  $j$. Thus if we simply want to rank the columns of DS1 according to
  their ``similarity'' to $C'_j$, we can simply compute the ratios
\begin{equation}
\label{eq:ag-cos3}
s_{ij} =
 \frac{ R(C_i,C'_j) }{ R(C_i,C_i)^{1/2} }
\end{equation}
When the matrix of these ratios $s_{ij}$ is reported as the assignment
matrix one can compare values within the same row of the matrix, but
not between rows.

\section{Unequal-size Samples from Different Columns} \label{sec:unequal}


The preceding discussion assumes that we have
data for an equal number ($N_1$) of cells from each column of
DS1. Similarly, we had $N_2$ cells from each column of DS2.

What if we have differently-sized samples from different columns of a
data source? This situation might result from the sampling process, or
might flow from a decision to {\em ignore} empty cells of the
data source, instead of choosing to treat them as legitimate cells
containing a value NUL.

Now, our sample of DS1's column $C_i$ will consist of $n_i$
cells; $N_1$ will be understood as $\max_i n_i$. Similarly, column
$C'_j$ of DS2 will have $n'_j$ cells, and $N_2=\max_j n'_j$.

How will this situation affect the formulas for the assignment matrix
elements proposed above?

It appears that the formulae for the {\bf arithmetic mean} (\ref{eq:ag-ar})
and {\bf geometric mean} (\ref{eq:ag-geo}) of the alignment values don't need
to be modified. Note that, by the definition of Bayesian probabilities, if
a particular (sampled) column $C_l$ of DS1 has exactly the same
composition of values of the column $C_i$, but we have fewer cells in
our samples from $C_l$ than from $C_i$ (i.e., $n_l < n_i$, then all
assignment elements
 $f(C_l|V)$ will be proportionally smaller than
$f(C_i|V)$:
$$
\frac{f(C_l|V)}{f(C_i|V)} = \frac{n_l}{n_i}.
$$ The arithmetic and geometric averages $f_{lj}$ will, too, be
proportionally smaller than $f_{ij}$, i.e. $f_{lj}/f_{ij}=n_l/n_i$.

To extend the {\bf weighted cosine similarity} formula
(\ref{eq:ag-cos2}) and (\ref{eq:ag-cos3}) to the case of unequal column samples,
 we can retain the original approach. We continue
defining $\alpha_i(V)$ as the fraction of the cells whose value is $V$
among the $n_i$ cells of column $C_i$. The values of $\gamma_j(V)$
will be defined similarly with respect to $C'_j$. We will still
define the cosine similarity $f_{ij}^{\rm sym 1}$ as the
cosine of the angle (in the weighted-dot-product space) between the
vectors $\vec{\alpha_i}$ and $\vec{\gamma_j}$; that is, if our samples
of columns $C_i$ and $C_l$ have exactly the same composition, even
though $n_i \ne n_l$, we'll want $f_{lj}^{\rm sym 1} =
f_{ij}^{\rm sym 1}$ for any $C'_j$.

With the above guidelines in mind, we note that, with a perfect Bayesian model,
$$f(C_i|V)=\alpha_i(V) n_i / (\sum_k \alpha_k(V) n_k),
$$
and
\begin{equation}
\label{eq:R-def2}
R(C_i,C'_j)\equiv \frac{1}{n'_j}\sum_{l=1}^{n'_j} P(C_i|z'_lj)  =
n_i \sum_V \frac{\alpha_i(V)\gamma_j(V)}{\sum_k \alpha_k(V) n_k}.
\end{equation}
We can thus define the weights
$$
\phi(V) =  \frac{1}{\sum_k \alpha_k(V) n_k}
$$ for use in our dot product, and express dot products in terms of
model alignment elements, $R(C,C')$
$$
(\vec{\alpha_i}, \vec{\gamma_j}) = R(C_i,C'_j)/n_i,
$$
$$
(\vec{\alpha_i}, \vec{\alpha_i}) = R(C_i,C_i)/n_i.
$$
This gives us the following generalization for (\ref{eq:ag-cos2}):
\begin{equation}
\label{eq:ag-cos-gen-2}
f_{ij}^{\rm sym 1} =
\frac{1}{\sqrt{n_i}} \cdot
\frac{ R(C_i,C'_j) }{ R(C_i,C_i)^{1/2} \|\vec{\gamma_j}\|}
\end{equation}
where, however,
$$
\|\vec{\gamma_j}\| \equiv \left(\sum_{v\in \cal{V}}{\gamma_j(v)^2  \phi(v)}\right)^{1/2}
$$ is not expressible in general probability terms. As in the
case of equal samples,  we note that the value
$\|\vec{\gamma_j}\|$ is the same for all elements in the same row of
the assignment matrix. We thus can generalize eq. (\ref{eq:ag-cos3}) as
\begin{equation}
\label{eq:ag-cos-gen-3}
s_{ij} = \frac{N_1}{\sqrt{n_i}} \cdot
 \frac{ R(C_i,C'_j) }{ R(C_i,C_i)^{1/2} }
\end{equation}
As with (\ref{eq:ag-cos3}), when the matrix of these ratios $s_{ij}$
is reported as the assignment matrix, one can compare values within the
same row of this matrix, but not between rows.

\section{A Symmetric-cosine Approach (``Symmetric No. 1'')}\footnote{Note that this is available in BOA with the {\tt sym1} command.} \label{sec:sym1}

Here we will propose an alternative approach to that outlined in
Sections \ref{sec:overview} and \ref{sec:agg}. While somewhat ``strange''
in its design, it will generate an assignment matrix with two pleasing
properties:
\begin{enumerate}
\item If the two data sources are identical, the matrix will be symmetric.
\item If the two columns $C_i$ and $C'_j$ are identical, the matrix element $f_ij$ will be equal to 1.
\end{enumerate}

The algorithm (outlined in the general, unequal-column-size, case) is
as follows:

\begin{enumerate}
\item[1.] Consider the set of $M_1+M_2$ fields of DS1 and DS2 as a single
  discrimination (set of labels) with $M_1+M_2$ classes (one per field)

\item[2.] Create $N_e = \sum_{i=1}^{M_1} n_i + \sum_{i=1}^{M_2} n'_i$
  training examples, each example being the content of one field of
  one record from DS1 or DS2, and carrying the class label based on
  the name of the data set combined with the name of the field in
  question. (When the data are represented with records as rows and
  fields as columns, each ``example" introduced at this step will
  correspond to the content of one cell of this table).

\item[3.] Tokenize etc. each ``example" somehow, converting it into a
  vector in some linear space (a feature vector)

\item[4.] Use some kind of Bayesian regression learning algorithm, such
  as one of those implemented by BOXER toolkit learner, on that set of $N_e$
  ``examples,'' to produce a classifier model that
  probabilistically assigns each ``example'' to a ``class'' (i.e., a
  field).

\item[5.] For each pair of columns from DS1+DS2, compute the assignment
  matrix value
\begin{equation}
\label{eq:ag-sym1}
f^{\rm sym 1} = \sqrt{\frac{ R(C_i|C'_j) R(C'_j|C_i) }{ R(C_i|C_i) R(C'_j|C'_j)}},
\end{equation}
where the averaged alignment values $R(C_i|C'_j)$ are computed as in (\ref{eq:R-def2}).
\end{enumerate}

It can be shown that the similarity value (\ref{eq:ag-sym1}) is the
cosine of the angle between the vectors $\vec\alpha_i$ and
$\vec\gamma_j$ in the Euclidean space where the dot product is defined
with the weight
$$
\phi(V) =
 \frac{1}{ \sum_{i=1}^{M_1} \alpha_i(V) n_i + \sum_{i=1}^{M_2} \gamma_i(V) n'_i}.
$$

{\bf Criticism.} This approach seems sensible when Assumptions 1 and 2
hold (i.e., both data source cells are from the same limited
vocabulary). However, it may have an unpleasant drawback in a
situation where free-form texts are stored in some columns.  Since the
learner is allowed to train on the combined discrimination including
cells from the columns of both data sources, it is rewarded for
finding features that distinguish columns from the two data sources
that otherwise would be viewed as fundamentally similar. For example, if texts in the
cells of column $C_a$ of DS1 are, overall, fairly similar to those in
the cells of column $C'_b$ of DS2, but contain some unique ''shibboleth''
word not found in DS2, then a well-trained learner {\em will} make
use of that word to construct a model that views $C_a$ as completely
distinct from $C'_b$.

\section{A Second Symmetric-cosine Approach (``Symmetric No. 2'')}\footnote{This is available in BOA with the {\tt sym2} command.}\label{sec:sym2}

The approach outlined in this section does not have much of a
theoretical foundation at present, but is also symmetric in
the sense that if an ontology is matched against itself, a symmetric
assignment matrix is produced.

\begin{enumerate}
\item[1.] Construct a PLRM model based on the cells of DS1, exactly as
  outlined in Section \ref{sec:overview}, Steps 1-4. In this section,
  $f(C_i|\cdot)$ and $R(C_i, \cdot)$ will refer to the alignments
   to the columns of DS1, and their aggregates (defined
  as per eq. (\ref{eq:R-def2})) obtained by this DS1-based model.

\item[2.] Construct another PLRM model based on the cells of DS2, in a
  similar way. The notation $P'(C'_i|\cdot)$ and $R'(C'_i, \cdot)$ will
  refer to alignments and their aggregates obtained by
  this DS2-based model.

\item[3.] Apply the first model to the cells of DS2, and the second
  model, to the cell of DS1. Compute the alignment matrix elements as
  follows:
\begin{equation}
\label{eq:ag-sym2}
f^{\rm sym 2}_{ij} = \sqrt{ \frac{ R(C_i|C'_j) R'(C'_j|C_i) }{ R(C_i|C_i) R'(C'_j|C'_j)}}
\end{equation}
\end{enumerate}

\section{Representative Applications}

We have applied these models to a test case produced by working with a
single large data collection. This collection is divided into two
parts, and we assess the algorithms for their ability to match each
field to itself, when it appears as an unlabeled test case. The
specific collection is the publicly available portion of the WITS data set.

The WITS data were maintained by the National Counterterrorism Center and provided a public interface to a substantial amount of anonymized data about terrorism incidents around the world. While the {\em underlying data} are perhaps very rich, the removal of proper names makes the set with which we conducted these experiments rather less interesting. Nonetheless, we find that some systematic trends appear when we examine both the effects of system parameters, and the nature of the cases in which the system succeeds or fails. In 2010 it was merged into the University of Maryland DHS Center of Excellence, START data based, which is available at: https://www.start.umd.edu/gtd/
An excellent discussion of the database is provided by Wigle (\cite{PoT88}). 

As noted, the contents of the working data set are  quite impoverished, which makes any sort of alignment difficult. For example, the 15th field contains information about the target of terrorist attack. The most frequent entries in this field are as shown below, and a substantial fraction of the records have no entry in this field.

\begin{verbatim}
    [sort WITS_15.txt | uniq -c | sort -nr ]
    451 NUL
    165 Vehicle
     49 Residence
     44 Community
     33 Transportation Infrastructure
     33 Energy Infrastructure
     31 Police
     29 Government
     26 Public Place/Retail
     26 Bus
     24 Business
     23 Religious
     11 Communications
     11 Checkpoint

\end{verbatim}

The WITS data also uses additional fields to report on second and third labels to be assigned as part of the same information. This might occur because there are multiple targets, or because a single target can be described in more than one way. [For example, a retail business with the proprietor's residence on the second floor.] Not surprisingly, additional fields, such as the third, shown here, are much sparser, and contain little semantic information.

\begin{verbatim}
sort WITS_17.txt | uniq -c | sort -nr
    987 NUL
      6 Vehicle
      6 Residence
      2 Military
      1 Unknown
      1 Religious
      1 Public Place/Retail
      1 Hospital/Health Care
      1 Government
      1 FacilityList_Facility_FacilityType3
      1 Diplomatic
      1 Community
      1 Communications
      1 Business
      1 Bus

\end{verbatim}

Another field, containing the reported nationality of the first victim of an attack is filled in a majority of the cases, and has a very long tail, containing 40 distinct countries.  The distribution of words, or of {\em n-grams} in such a   field will be unstable over time (say, on the scale of months) as the incidence of terrorist attacks, and the toll that they take varies over time. As an example, the toll in Israel was higher in 2006 that it would be for a corresponding month in 2011, while that for Pakistan was lower five years ago. Nonetheless, over short time periods this may be stable enough to permit alignment. This shows that international data, which will draw from many different lexicons, even when all translated into English, can pose challenges to the BOA approach. We anticipate that it will be more effective within a single country, such as the United States. Even so, we would expect that, for example, the distribution of surnmes will be quite different in, for example, Miami and Detroit.

DECIDE WHETHER TO PUT THIS ON THE WEB. It is on Paul's computer at home. 
\newpage

To get an intuitive feeling for the effectiveness of an alignment process, we show a portion of the very large alignment matrix found using the BOA  Stochastic Gradient Descent (sgd) option, with the specific rather fast shrinkage: $\eta=0.01$ and the number of passes through the data at a rather large number: $rep=2,000$. The extracts are shown in Tables \ref{tab:extract1} and \ref{tab:extract2}.


\begin{table}
\begin{center}
\begin{tabularx}{.7\textwidth}{>{\bfseries}l|c c c c c c c c|}
  \hhline{~--------}
  	& (1)&	(2)&	(3)&	(4)&	(5)&	(6)&	(7)&	(8)\\
   \hhline{~--------}
(1) ICN&	{\bf 99.9} \cellcolor[gray]{.8} &	0.0&	0.0&	0.0&	0.0&	 0.0&	0.0&	 0.0\\
(2) IED&	0.0&	{\bf 41.2} \cellcolor[gray]{.8} &	0.0&	0.0&	0.0&	 0.0&	0.0&	 0.0\\
(3) Incident Date&	0.7&	0.0&	{\bf 98.8} \cellcolor[gray]{.8} &	0.0&	 0.0&	0.0&	0.0&	 0.0\\
(4) City1&	0.0&	0.0&	0.0&	{\bf 81.6} \cellcolor[gray]{.8} &	1.7 \cellcolor[gray]{.9}&	 14.8&	1.7&	 2.1\\
(5) City2&	0.0&	0.0&	0.0&	0.1&	 13.0 \cellcolor[gray]{.8} &	 0.0&	13.3&	0.0\\
(6) State Province1&	0.0&	0.0&	0.0&	18.8&	0.1&	{\bf 84.9} \cellcolor[gray]{.8} &	 2.3 \cellcolor[gray]{.9} &	 0.2\\
(7) State Province2&	0.0&	0.0&	0.0&	0.1&	13.0&	0.0&	13.3 \cellcolor[gray]{.8}&	 0.0\\
(8) Country&	0.0&	0.0&	0.0&	4.2&	0.0&	0.1&	0.0&	55.7 \cellcolor[gray]{.8}\\

  \hhline{~--------}
\end{tabularx}
\caption{An extract from the alignment matrix with 89 training columns and 153  rows to be labeled. The entry in each cell is the alignment number for the corresponding row to be matched to the corresponding column. Bold font indicates that this column is the best match for the given row, the darker grey indicates the exact match to the field of the same name. A lighter gray indicates that the match is to an auxiliary field of the same type (such as the second victim). As is clear, in all but one case, the best match is also the correct one. The failure, that is, the row without a bold entry,  is an example of the cases where distributional information cannot be adequate. The target row is an auxiliary list, and it cannot be matched exactly to the {\em corresponding  auxiliary list with the same kinds of entries}. Thus the country in which the event occurs was automatically matched to a field containing the nationality of the victims, which is not shown here.  This is of course a ``reasonable sort of error.'' }
\label{tab:extract1}
\end{center}
\end{table}

\begin{table}
\begin{center}
\begin{tabularx}{.7\textwidth}{>{\bfseries}l|c c c c c c c c c|}
&	(9)&	(10)&	(11)&	(12)&	(13)&	(14)&	(15)&	(16)&	(17)\\
\hhline{~---------}
(9) Region&	{\bf 100 }  \cellcolor[gray]{.8}&	0.0&	0.0&	0.0&	0.0&	 0.0&	0.0&	0.0&	 0.0\\
(10) Multiple Days&	0.0&	37.0  \cellcolor[gray]{.8}&	0.0&	0.0&	0.0&	 0.0&	0.0&	{\bf 37.3}&	0.0\\
(11) Characteristic1&	0.0&	0.0&	{\bf 58.9}  \cellcolor[gray]{.8}&	 0.4  \cellcolor[gray]{.9}&	26.8&	0.1&	 0.0&	 0.0&	0.0\\
(12) Characteristic2&	0.0&	0.0&	5.3 \cellcolor[gray]{.9}&	12.9  \cellcolor[gray]{.8}&	0.0&	 12.9&	0.0&	 0.0&	 0.0\\
(13) Nationality1&	0.0&	0.0&	16.7&	0.1&	{\bf 40.2}  \cellcolor[gray]{.8}&	0.5  \cellcolor[gray]{.9}&	 0.0&	 0.0&	0.0\\
(14) Nationality2&	0.0&	0.0&	0.0&	12.9&	0.7 \cellcolor[gray]{.9}&	 12.9  \cellcolor[gray]{.8}&	0.0&	 0.0&	 0.0\\
(15) Subject&	0.0&	0.0&	0.0&	0.0&	0.0&	0.0&	{\bf 97.0}  \cellcolor[gray]{.8}&	 0.0&	0.0\\
(16) Suicide&	0.0&	36.3&	0.0&	0.0&	0.0&	0.0&	0.0&	{\bf 36.7} \cellcolor[gray]{.8}&	0.0\\
(17) Summary&	0.0&	0.0&	0.0&	0.0&	0.0&	0.0&	0.0&	0.0&	 {\bf 99.3} \cellcolor[gray]{.8}\\
 \hhline{~---------}
\end{tabularx}
\caption{The diagonal portion of the alignment matrix, continued. For clarity of presentation, none of the off-diagonal blocks of this portion of the array are shown. They do not contain any features of interest (that is, false alignments).}
\label{tab:extract2}
\end{center}
\end{table}

\begin{figure}
\begin{center}
\includegraphics[width=200mm]{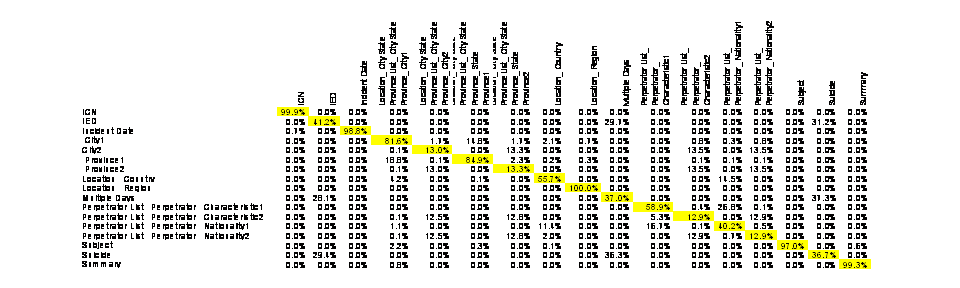}
\end{center}
\caption{A visual image of the alignment matrix for the key information from a selection of WITS records, the details are shown in Tables \ref{tab:extract1} and \ref{tab:extract2}}.
\label{fig:extract}
\end{figure}

A visual image of the larger array, which is unfortunately too small to read easily, is shown in Figure \ref{fig:extract}.

Another field (number 46) contains rich unstructured information, called the {\bf Summary}. At the request of the sponsor, these summaries are not included in this report. Some examples, suitably minimized are shown here. If numerous such examples contain the same geographic place names (represented here as [GEO]) that supports a very accurate alignment. Alignment is further aided by the use of standard phrases such as ``assailant.'' This is an artifact of the example case, in which a single base has been divided to illustrate the method. Another database might, for example, have used the terms ``suspect,'' or ''alleged perpetrator.''

\begin{em}
\begin{quote}

WITS\_46.txt:On 5 March 2006, in [GEO]  District, [GEO], [GEO] , assailants on a motorcycle fired upon and killed a [GROUP]  civilian as he drove on a deserted road.  No group claimed responsibility, but it was widely believed that [GROUP]   were responsible.

WITS\_46.txt:On 6 March 2006, early in the day, in [GEO] Village, [GEO] , [GEO], [GEO], a large group of assailants stormed a [GROUP] village and fired upon two houses, killing three civilians, injuring one other, burning the body of one victim, and setting fire to two houses.   No group claimed responsibility, but authorities stated that [GROUP]separatists were responsible.

\end{quote}
\end{em}

\section{Evaluating the Algorithms}

To be able to summarize the quality of the alignment matrix in just a
few numbers we used the following technique, which is only applicable
when the ``correct'' correspondence between the two data sources is
known, in particular when the two data sources have the same number of
columns $M=M_1=M_2$.

We can ask: In how many of the $M$ rows of the
alignment matrix the ``correct'' column contains the largest value of
this row (as it ought to be, in case of the correct matching)? In how
many rows the ``correct'' column contains one of the two largest
values of the row? One of the three largest values? (Actually, we did
not ask about ``largest,'' ``second largest,'' etc., value, but
``ranked-1,'' ``ranked-2,'' etc., value, with the ranking using a somewhat
arbitrary tie-breaking procedure when the actual values of several
matrix elements were identical). The perfect matching, of course,
would have correctly put each row's top value into the correct column.

For one series of experiments we used the data set
{\tt   safeWITS\_006\_03}
  , with $M=89$ fields.  From one large table
with 89 fields we have selected the first 100 entries as a training set, and the {\em same} group
of 100 entries,

The simplest task is to use the same first group as both DS1 and DS2
(this is referred to as ``same section,'' in the table below). The
algorithm's performance on a task like this is not, of course,
indicative of its ability to match different data; it simply measures
the degree to which the model is able to (over-)fit the data set.

A somewhat more complex task is to match the same DS1 against a DS2
selected from a different, but closely related data source: a 100-row
excerpt (rows 500 through 599) from \newline {\tt
  Sent2010.12.21WITS\_2006\_04}.  This seeks to model April of 2006,
in terms of March 2006.

In this second experiment, the latter table has $M_2=153$ columns; however, an
$M_1$-sized subset of the columns of DS2 does have a direct
correspondence to all columns of DS1. The remaining columns of DS2
were what one can call ``overflow columns,'' intended to contain data
of the same type that other columns (present also in DS1) contain when, for
example, one needs to store a list of multiple place names instead of
a single place name. See Table \ref{153to89}.

The results in the table below have been obtained as follows. For features, we used words and n-grams up to $n=2$, plus the special ``empty cell'' feature. For the second step, we used a process that provides a close approximation to the maximum log-likelihood (Adaptive Steepest Descent with a low value for the parameter $\epsilon$).


\subsection{PLRM classifier obtained with $\epsilon=10^{-8}$:}

\begin{table}[ht]
\caption{Matching 89 fields of WITS data to themselves or another month}
\begin{center}
\begin{tabular}{|l | r| r|}
\hline
Aggregation method & Same source &  Different source \\
\hline
Avg. arith. & 34/50/55 & 24/34/39 \\
Avg. geom. & 65/76/81 & 43/56/62 \\
Cosine & 72/79/81 &  34/42/47 \\
\hline
Sym1 & 81/86/87 & 32/45/47 \\
Sym2 & 82/86/87 & 35/47/54 \\
\hline
\end{tabular}
\end{center}
\title{\em  In each cell the three numbers x/y/z indicate the specific number of target fields that are correctly aligned to their definitions when examining only the top ranked field name (x), or the top 2 (y), or the top 3 (z). It is felt that a human user interacting with the system might be willing to look at no more than three candidates before resorting to a more sophisticated ontology alignment tool. The symmetric methods bring 87 of 89 [almost 98\%] into one of the top three positions. The same 89 fields are considered, using either the same data as the test set (same) or a different 100 rows of data from the same month (different). }
\label{tab:scoring1}
\end{table}


\subsection{$k$-NN classfier with $k=3$}
For comparison with a non-Bayesian method we show the results of solving the same problem using a $kNN$ method with the number of near neighbors set to $3$. Such non-parametric methods do not require any assumption of linearity in the feature space. As a practical matter, if non-parametric methods perform dramatically better, it suggests that the linear classifier approach used here would not be effective.  Such methods do have substantial challenges in very high dimensions, although there are contemporary non-linear developments in manifold reduction that provide an alternative approach.

\begin{table}[ht]
\caption{Matching 153 fields of WITS data to 89 basic fields}
\begin{center}
\begin{tabular}{|l | r| r| }
\hline
Aggregation method & Same section & Different source \\
\hline
Avg. arith. & 34/50/55 & 24/34/39 \\
Avg. geom. & 65/77/82 & 26/34/36 \\
Cosine & 72/79/81 & 33/42/47 \\
\hline
\end{tabular}
\label{153to89}
\end{center}
\title{\em The notation is as in Table 3,  but the set of fields to be classified now contains multiple fields with the same kind of data (e.g., ``second victim''), which is not different in content, but will be different in sparsity from the primary field. So multiple fields can properly match to the same ``ontological'' field of the model.  }
\end{table}

\section{Experimenting with Different Representation Schemes}

The following shorthand notation is used for representation scheme names: {\tt e*-w*-g*}. The meaning of the elements is as follows:
\begin{itemize}
\item {\tt e0} that special NUL tokens for empty cells are {\em not} used
\item {\tt e1}  that special NUL tokens for empty cells are  used
\item {\tt w0} {\em no} features for words
\item {\tt w1}   features for words
\item {\tt g0} {\em no} features for $n$-grams
\item {\tt g1} features for 1-grams (i.e., characters)
\item {\tt g2} features for 1- and 2-grams (i.e., characters and 2-grams)
\item {\tt g3} features for 1-, 2-, and 3-grams
\item etc.
\end{itemize}

In the tables in this section, ``same source'' means that the two data sources were identical (same section of the same file); ``different source'' means that the second data source contained data from a different month from a file with a different (larger) scheme, as in the tables shown above.

\subsection{Words, (1,2)-grams, and special NUL token}

(e1-w1-g2) (Empty cell=special token, words=true, grams=(1,2))

The features are words, characters (``1-grams''), 2-grams, and the special token for empty cells.

{\footnotesize
\begin{table}[ht]
\caption{Words, (1,2)-grams, and the NUL token}
\title{Number of columns in the target table that are assigned correctly by the
highest/first two/first three selections in the alignment matrix. Note that for assignment from a different month, and for a larger number of target columns to be labeled, the performance is not as good, as discussed above. }
\begin{tabular}{|l | r| r|r| r|r| r|r| r|}
\hline
 & \multicolumn{8}{|c|}{\bf $\epsilon$ parameter} \\
\hline
Aggreg. & $10^{-1}$ & $10^{-2}$ & $10^{-3}$ & $10^{-4}$ & $10^{-5}$ & $10^{-6}$ & $10^{-7}$ & $10^{-8}$ \\
\hline
Same src:  \\
\hline
Avg. arith.& 1/4/5 & 7/12/15       & 24/29/38      & 25/39/44      & 31/47/51      & 34/49/54      & 34/50/55      & 34/50/55  \\
Avg. geom. & 1/4/5 & 7/11/14       & 33/45/48      & 51/58/61      & 66/75/78      & 67/75/80      & 66/76/80      & 65/76/81  \\
Cosine     & 16/26/31 & 25/36/41   & 37/51/57      & 46/58/61      & 69/73/74      & 73/77/78      & 74/78/79      & 72/79/81 \\
\hline
Diff. src:   \\
\hline
Avg. arith.& 1/4/5 & 9/13/16      & 21/29/34      & 23/33/36      & 23/34/39      & 24/35/39      & 24/35/39      & 24/34/39 \\
Avg. geom.& 1/4/5 & 9/13/17       & 23/31/35      & 35/45/50      & 43/54/61      & 43/55/63      & 44/56/64      & 43/56/62 \\
Cosine    & 15/26/33  & 23/33/37  & 28/40/47      & 28/40/50      & 33/49/53      & 33/42/47      & 33/42/47      & 34/42/47  \\
\hline
\end{tabular}
\title{\em  In each cell the three numbers x/y/z indicate the specific number of target fields that are correctly aligned to their definitions when examining only the top ranked field name (x), or the top 2 (y), or the top 3 (z). It is felt that a human user interacting with the system might be willing to look at no more than three candidates before resorting to a more sophisticated ontology alignment tool.  }
\end{table}
}

\subsection{Words and (1,2)-grams}

(e0-w1-g2) {Empty cell=zero vector, words=true, grams=(1,2)}

Here our features are words, characters, and 2-grams. No
special NUL token for empty cells is used.

{\footnotesize
\begin{table}[ht]
\caption{Words and (1,2)-grams}
\begin{tabular}{|l | r| r|r| r|r| r|r| r|}
\hline
 & \multicolumn{8}{|c|}{\bf $\epsilon$ parameter} \\
\hline
Aggreg. & $10^{-1}$ & $10^{-2}$ & $10^{-3}$ & $10^{-4}$ & $10^{-5}$ & $10^{-6}$ & $10^{-7}$ & $10^{-8}$ \\
\hline
Same src:  \\
\hline
Avg. arith.& 1/3/5 & 2/5/5 & 21/27/35      & 24/38/42      & 26/40/46      & 35/49/53      & 34/50/55      & 34/50/55\\
Avg. geom. & 1/3/5 & 2/5/5 & 26/37/44      & 48/60/61      & 58/70/71      & 66/76/81      & 67/76/80      & 66/76/80\\
Cosine & 17/21/24  & 21/25/28   & 26/37/48 & 52/60/62      & 64/67/72      & 69/77/78      & 71/76/78      & 73/79/81\\
\hline
Diff. src:   \\
\hline
Avg. arith.& 1/3/5 & 2/4/5 & 20/29/33      & 23/33/39      & 23/34/40      & 24/33/38      & 24/35/39      & 24/34/39\\
Avg. geom.& 1/3/5 & 2/4/5 & 24/36/43      & 31/43/49      & 43/51/57      & 44/53/60      & 44/56/65      & 43/56/63\\
Cosine& 16/20/22 & 18/25/29 & 22/34/39      & 29/44/51      & 33/45/50      & 34/48/53      & 33/45/50      & 33/42/47\\
\hline
\end{tabular}
\title{\em  In each cell the three numbers x/y/z indicate the specific number of target fields that are correctly aligned to their definitions when examining only the top ranked field name (x), or the top 2 (y), or the top 3 (z). It is felt that a human user interacting with the system might be willing to look at no more than three candidates before resorting to a more sophisticated ontology alignment tool.  }
\end{table}
}

\subsection{Only words}

(e0-w1-g0)  (Empty cell=zero vector, words=true, grams=0)

Here, only words are used as features. No character or n-gram features, and no NUL token.

{\footnotesize
\begin{table}[ht]
\caption{Words only}
\begin{tabular}{|l | r| r|r| r|r| r|r| r|}
\hline
 & \multicolumn{8}{|c|}{\bf $\epsilon$ parameter} \\
\hline
Aggreg. & $10^{-1}$ & $10^{-2}$ & $10^{-3}$ & $10^{-4}$ & $10^{-5}$ & $10^{-6}$ & $10^{-7}$ & $10^{-8}$ \\
\hline
\multicolumn{9}{|c|}{\bf Same Source} \\ \hline
Avg. arith.	& 4/7/8	& 22/32/37	& 25/37/42	& 27/40/44	& 27/40/45	& 27/40/45	 & 27/40/45	& 27/40/45 \\
Avg. geom. 	& 4/7/8	& 25/34/38	& 27/37/44	& 27/38/45	& 30/41/47	& 32/42/47	 & 32/42/47	& 35/42/49 \\
Cosine  	& 32/44/54	& 41/52/56	& 59/64/69	& 62/72/77	& 65/74/75	& 65/74/75	 & 65/74/75	& 65/74/75 \\
\hline

\multicolumn{9}{|c|}{\bf Different Source} \\ \hline
Avg. arith.	& 4/6/7	& 21/28/31	& 21/33/36	& 22/32/36	& 22/32/36	& 22/32/36	 & 22/32/36	& 22/31/36 \\
Avg. geom. 	& 4/6/7	& 20/29/31	& 22/31/35	& 23/31/38	& 23/33/39	& 23/33/38	 & 24/33/39	& 23/33/39 \\
Cosine  	& 24/35/47	& 26/38/46	& 33/44/52	& 33/41/54	& 32/43/53	& 31/43/53	 & 31/43/53	& 31/43/53 \\
\hline
\end{tabular}
\title{\em  In each cell the three numbers x/y/z indicate the specific number of target fields that are correctly aligned to their definitions when examining only the top ranked field name (x), or the top 2 (y), or the top 3 (z). It is felt that a human user interacting with the system might be willing to look at no more than three candidates before resorting to a more sophisticated ontology alignment tool.}
\end{table}
}


\subsection{e1-w1-g0}

(e1-w1-g0) (special NULL token, words, no n-grams)

Our features are words and the special NUL token for empty cells. No $n$-grams.

{\footnotesize
\begin{table}[!h]
\caption{Words and the special NULL token}
\begin{tabular}{|l | r| r|r| r|r| r|r| r|}
\hline
 & \multicolumn{8}{|c|}{\bf $\epsilon$ parameter} \\ \hline
Aggreg. & $10^{-1}$ & $10^{-2}$ & $10^{-3}$ & $10^{-4}$ & $10^{-5}$ & $10^{-6}$ & $10^{-7}$ & $10^{-8}$ \\
\hline
\multicolumn{9}{|c|}{\bf Same Source} \\ \hline
Avg. arith.	& 5/8/10	& 24/33/39	& 25/39/42	& 33/47/52	& 34/50/54	& 34/50/55	 & 34/50/55	& 34/50/55 \\
Avg. geom. 	& 5/8/10	& 43/50/55	& 54/60/62	& 63/72/75	& 67/76/80	& 66/76/80	 & 65/77/82	& 65/77/82 \\
Cosine  	& 18/25/33	& 40/51/53	& 53/55/58	& 63/68/72	& 74/78/79	& 74/78/79	 & 72/79/81	& 72/79/81 \\
\hline
\multicolumn{9}{|c|}{\bf Different Source} \\ \hline
Avg. arith.	& 6/8/10	& 22/28/32	& 22/33/37	& 23/33/37	& 23/34/38	& 23/34/38	 & 23/33/38	& 23/33/38 \\
Avg. geom. 	& 6/8/10	& 33/42/45	& 38/47/53	& 42/53/63	& 42/55/64	& 42/53/61	 & 43/53/61	& 43/54/61 \\
Cosine  	& 16/24/29	& 25/38/45	& 34/43/48	& 32/43/48	& 32/41/47	& 32/41/47	 & 32/41/47	& 32/41/47 \\
\hline
\end{tabular}
\title{\em Performance of various aggregation methods when tested on the same data on which is was built, or on held out data from the same month. The parameter $\epsilon$   controls the precision of the model. Decreasing $\epsilon$ increases the time required, but produces better results.}
\end{table}
}

The ``drop" in performance when going from the labeled data (DS1) to the
 unlabeled data (DS2) is not unexpected. Generally,  the symmetric methods are
designed to produce near-perfect results precisely when DS1=DS2.
When we test that, we obtain performance figures close to 89/89/89
and miss only because there are occasional exact ties. But all  of  this, of course, falls apart when one
deals with even slightly less than ideal data, e.g., a  DS2 that is  similar,
but not identical, to DS1.

\section{Directions for Future Research}

\subsection{Estimating confidence in the results}

If the BOA approach is to be used in practice, we should replace an informal estimate of user behavior with a computational estimate of accuracy. To do this, two steps are needed. The first is to interpret the alignment results in terms that naturally support a probabilistic interpretation. It appears that there are a number of different ways in which this might be done, as detailed below.

\subsection{Theory-driven estimation}

For example, with the ``average arithmetic" aggregation methods, the
numbers in each column do sum to 1.0, just as probabilities
would. Whether it is wise to think of them as probabilities is
another question.  In addition, these are not the most effective of the methods.

Generally, one can always transform each row of the ``alignment matrix" so
that all numbers in each row will sum to 1.0. (For example, simply
by normalizing). However, this is so artificial that it can hardly be
thought of as a valid probability interpretation.

We enumerate here a number of possible approaches
in the context of our ``forward" technique, whose description
occupies most of this paper. Since the ``symmetric''
methods appear more effective, future work will explore extensions of these ideas.

(1) Even at the stage of generating the classifier
for the cells of DS1 we can ask ``how good is the classifier'' (when
applied to the training set itself). We can easily answer that
question in terms of the classifier's ``lin-likelihood" (the average
probability  estimate for the assignment of cells of DS1 to their correct
columns) or its log-likelihood (the average of the logarithms of those probabilities). This kind of estimate conveys an
assessment
of the extent to which  individual cells of DS1 can be unambiguously
recognized as belonging to their particular columns. In other words, if the table were
printed on paper, with each cell's content written in the appropriate
square, and then cut into squares - can the classifier put all the
squares into their original columns again?

The upper bound (1.0 for lin-lik, or 0.0 for log-lik) is [nearly]
achieved when all elements can be unambiguously assigned to columns.
If the same value (e.g., ``Yes'' or an empty value, NUL) occurs in multiple
columns, then of course neither of these measures can reach the upper bound.
More generally, the bound cannot be achieved if the texts in
different columns' cells are distinct, but not linearly separable
within the feature framework being applied (such as {\em 4-grams}).

(2) How does this measure for the quality of the classifier relate to its
performance on DS2? It seems obvious enough that if the classifier
itself is poor, then its results on DS2 should not be given much
credence either. Of course,  theoretically, one can imagine a DS2 containing
only cells with ``easy to classify'' values and none of the
``harder'' (more ambiguous) values found in DS1.

But,  as is well known,
great performance on the training
set is no guarantee of performance on the test set, DS2,  either because
DS2 contains values not seen in DS1, yielding ``ambiguous'' scores
based on their features, or because values that the classifier can
identify well with a particular column of DS1 are spread over
multiple columns of DS2.

(3) Can BOA be made aware that it performs poorly on DS2? Of course
if it consistently assign all values from column DS2.x to some column DS1.a,
while human experts think that DS2.x really ought to match DS1.b, it
will never know that. But at least it can be made to assess its
confidence in the results it produces. Several ways to measure such
confidence can be proposed within our multi-step framework: both at
the level of the final ``alignment matrix," and at the level of the
individual scores for the cells of DS2.

(3a) At the level of the individual cell scores, we can compute
something like the lin-lik or log-lik over the cells from each column
of DS2. Of course to compute lin/log-lik one needs to know the
``correct class," but since all we want is to assess BOA's confidence
in its own results, it makes sense to take the DS1's column that BOA
currently proposes as the ``most likely match" for the column DS2.a of
DS2 for the notional ``correct class" for the cells of DS2.a.

(Incidentally, the alignment matrix element (DS2.a, DS1.x) for our
``avg geom'' and ``avg arith'' aggregation methods is based exactly on
the
log-lik and lin-lik, respectively, that we would obtain if we use the
cells of the column DS2.a as the training set, and the class DS1.x as
the ``correct class" for all of them.)

Our approach for interpreting the
alignment matrix is as follows: we choose the largest element in each
row of the matrix, and say that this gives the best match
for the given field of DS2. In other words, we match the field DS2.a
of DS2 to that field DS1.x which maximizes the alignment matrix
element (DS2.a, DS1.x). Depending on the aggregation method, this is
exactly the maximization of such ``local'' (field-wise) log- or
lin-likelihood. {\em Since this is local maximization, the same field of
DS1 may be chosen as the best match for several different fields of
DS2.} If this is undesirable, and we want to produce a 1-to-1
matching,
we can (at least, in the case of the datasets  having the same number of
fields) interpret the alignment matrix slightly differently: we can,
for whatever matrix we have obtained by the ``avg geom" /
``avg arith" aggregation, do a matching optimization, finding the permutation matrix that
maximizes the product or sum, respectively, of the alignment matrix
elements it selects. This will correspond to the global maximization
of log/lin-lik. This can be formulated as a bi-partite graph matching problem.

(3b) At the level of the final ``lignment matrix'' H,  it is
easiest, perhaps,  to devise a suitable measure for the avg arith
aggregation method. We can ask, ``what matrix Q containing only $0$ or $1$
with
exactly row sums of $1$ (or, with row and column sums $=1$, if a 1-to-1 mapping is desired) is closest to our
alignment matrix in terms of a 1-norm, $L_1$''? The 1-norm distance $|Q-H|$
will be our measure of confidence. (That is, $|Q-H|=0$ means that BOA
gave us a permutation matrix, or at least a matrix with one 1 per row
and 0s elsewhere - i.e., it is pretty confident! The maximum  possible
difference corresponds to the case when BOA gives equal values to all
possibilities, i.e., it knows that it does not know). Finding
this
Q, of course, corresponds to finding such matching of fields that it
maximizes the lin-likelihood (row-wise or global), much as mentioned
above.

Note that since we have avg arith aggregation, and each row of H
sums
to 1, it does not matter much whether we just look at the differences
between the elements of Q and H in the positions where Q has its
1s,
or in all positions; the two sums simply differ by a factor of 2.

In the case of the avg geom aggregation, one can try to do a similar
trick with logs of the matrix elements, but it appears we need only consider
 the values in the positions where H has its $1$s, in order
to
avoid dealing with infinities.

\subsection{Empirically driven estimation}

Complementing these approaches, one might also consider that the data in the alignment for DS1 offer an opportunity to ``calibrate'' the score against the probability of being correct. Whether this is feasible depends on whether, as an empirical matter, the threshold value for being (for example) right 80\% of the time has some consistent value across related applications of the BOA tool. This also requires further investigation with a variety of datasets.

\subsection{Possible Use Cases}

In this report, with a limited set of test data, we have used some assumptions to facilitate the presentation. These can removed without difficulty. They are 
{\bf Assumption 1.} The elements of $\cal{V_1} \subset \cal{V}$, the
set of all values of cells of DS1, have been converted to linearly
independent vectors and  
{\bf Assumption 2.} Every entry found in DS2 is also found somewhere in DS1. In general we do not need to require linear independence, nor do require that the intersection of the two sets of field entries is non-empty.

In the present (2019) environment, this approach is of interest because it has the potential to enable small organizations to begin to make their data available to other small organizations, or to fusion centers. We sketch here a possible mechanism for such sharing. 

The range of participating organizations should agree on a representation to be used. For example, the 1-gram, 2-gram, and cell length representation yield a small number of integers (using 128 ASCI characters, there are 128 1-grams and 16,384 bigrams, plus one the range of possible cell lengths, perhaps 100 or 200 more. Thus each participating agency could run frequency counting algorithms over each field of each data bases, and represent every field by 128+16,384+a few hundred integers. The vector  is, of course, quite sparse, as many bigrams do not occur, for any given field. These can be used, together with the methods explored here, to compute the degree of match of any field in one data set to all of the fields in the others. This provides an automated way to begin the search for corresponding information, and to support complete alignment of one data set with another. While we have examined the process using a particular and fairly sophisticated matching process, for practical development one may begin with some simpler measures of similarity. With the proposed approach, each data set is characterized by three frequency distributions, for 1-grams, 2-grams, and string length. Metrics such as the Jensen-Shannon metric and its extensions (\cite{osan2018monoparametric}), can be applied to find the similarity or distance between any pair of distributions over the same set..

\section{Acknowledgments and Disclaimer}
This research is supported in part by the Air Force Research Laboratory (AFRL). The findings and opinions presented in this report are those of the authors and any opinions, findings and conclusions or recommendations expressed in this material do not necessarily reflect the views of AFRL. We thank Dr. John Wullert and Dr. Allen MacIntosh of Telcordia Technologies Inc. for many helpful discussions. This work was substantially completed in 2012, but has been updated, in this version, to relate the topic to important current problems. One of us (PK) thanks the University of Wisconsin Department of Industrial and Systems Engineering for additional support. 

\newpage

\bibliographystyle{plainnat}
\bibliography{KDD,KDDBOA,BOXER-SIGIR-2010}

\end{document}